\documentclass[a4paper,UKenglish,cleveref, autoref, thm-restate]{lipics-v2021}
\newcommand{\rev}[1]{{\color{black}#1}}
\newcommand{\revb}[1]{{\color{black}#1}}
\newcommand{\revc}[1]{{\color{black}#1}}

\title{The Impact of Configuring Agentic AI Coding Tools on Build-vs-Buy Decisions: A Study Protocol}
\titlerunning{The Impact of Configuring Agentic AI Coding Tools on Build-vs-Buy Decisions}

\author{Jai Lal Lulla}{Singapore Management University, Singapore}{jailal.l.2025@phdcs.smu.edu.sg}{https://orcid.org/0009-0005-0024-8238}{}

\author{Matthias Galster}{University of Bamberg, Germany}{mgalster@ieee.org}{https://orcid.org/0000-0003-3491-1833}{}

\author{Jie M. Zhang}{King's College London, United Kingdom}{jie.zhang@kcl.ac.uk}{https://orcid.org/0000-0003-0481-7264}{}

\author{Sebastian Baltes}{Heidelberg University, Germany}{sebastian.baltes@uni-heidelberg.de}{https://orcid.org/0000-0002-2442-7522}{}

\author{Christoph Treude}{Singapore Management University, Singapore}{ctreude@smu.edu.sg}{https://orcid.org/0000-0002-6919-2149}{}

\authorrunning{Lulla et al.} 

\Copyright{Lulla et al.} 

\ccsdesc[500]{Software and its engineering~Software creation and management}
\ccsdesc[500]{Software and its engineering~Software notations and tools~Software configuration management and version control systems}
\ccsdesc[500]{Software and its engineering~Software notations and tools~Software libraries and repositories}
\ccsdesc[300]{Computing methodologies~Artificial intelligence}

\keywords{Agentic AI Coding Tools, Dependency Management, Build-vs-Buy, Configuration Mechanisms, Claude Code, OpenAI Codex, Software Engineering, Software Maintenance}

\category{Registered Report -- Confirmatory Track} 

\EventEditors{Bianca Trinkenreich and Marvin Wyrich}
\EventNoEds{2}
\EventLongTitle{Empirical Software Engineering International Week 2026 (ESEIW 2026), ESEM Registered Reports Track}
\EventShortTitle{ESEIW 2026}
\EventAcronym{ESEIW}
\EventYear{2026}
\EventDate{October 4--9, 2026}
\EventLocation{Munich, Germany}
\EventLogo{}
\SeriesVolume{XX}
\ArticleNo{XX}

\begin{document}
\nolinenumbers 

\maketitle

\begin{abstract}
Agentic AI coding tools write code with increasing autonomy and in doing so decide when to import a library and when to implement functionality from scratch. These decisions, whether to \emph{build} functionality from scratch or \emph{buy} into an external library, hereafter \emph{build-versus-buy}, carry direct consequences for software security, licensing compliance, performance, and long-term maintainability. Yet no controlled experimental study has examined what governs build-versus-buy decisions in agentic AI coding tools. Configuration mechanisms, i.e., the means by which developers tailor agentic AI coding tool behavior to a project or workflow, are one of the primary means by which practitioners can influence these decisions. However, it is unclear which configuration mechanisms influence build-versus-buy decisions most effectively. We present a pre-registered protocol to study how configuration mechanisms alter build-versus-buy behavior in two popular agentic AI coding tools: Claude Code and OpenAI Codex. We will execute controlled programming tasks drawn from a benchmark of staged projects, each constructed around identifiable build-versus-buy points, and will manipulate the configuration supplied to each tool, ranging from no configuration, through context files with soft preferences and explicit prohibitions, to Skills (instructions that can be autonomously discovered), MCP-enabled library discovery tools, and permission controls, measuring which libraries the tool selects, whether it discloses newly introduced libraries, and whether those disclosures are complete and accurate. Nine pre-registered hypotheses structure the protocol. \rev{We design the protocol for replication, so that the same measurement procedure can be re-executed against future agentic AI coding tools and model versions, supporting longitudinal study of how build-versus-buy decisions evolve as the underlying technology changes.} The resulting benchmark dataset \rev{and analysis pipeline} will be released as a reusable artifact for evaluating build-versus-buy behavior in agentic AI coding tools.
\end{abstract}

\section{Introduction}
\label{sec:intro}

Software dependencies are commitments. Each external library a project imports carries licensing obligations, exposes an attack surface, and creates a maintenance burden that compounds over time~\cite{zimmermann2019usenix}. The decision of whether to implement a capability internally (\emph{build}) or adopt an external library (\emph{buy}), the \emph{build-versus-buy} decision, used to belong to human engineers who weighed project constraints, organizational policy, performance requirements, and library quality before committing.

These decision dynamics have begun to shift. Agentic AI coding tools accept high-level task descriptions and plan, write, and execute code with minimal per-step human direction, and they introduce external dependencies without explicit instruction~\cite{anthropicclaudecode, openaicodex}. For example, when a developer asks such a tool to implement authentication, it may write \texttt{import bcrypt} or issue a \texttt{pip3 install passlib} command without stating so in its natural-language reply. The dependency enters the codebase unannounced. If the developer does not review code changes systematically, it may not be noticed until a security audit or license scan surfaces it.

\rev{This shift creates practical problems for practitioners. As adoption of these tools grows~\cite{robbes2026agenticmuchadoptioncoding}, observational data~\cite{ong2026claude} confirms that systematic library preferences are already visible in deployed releases.} An agentic AI coding tool that systematically favors certain libraries embeds those choices across many codebases, amplifying both the benefits and the risks of those libraries in ways no single developer decision would. Tools that do not disclose the dependencies they introduce undermine the human oversight that security and compliance reviews depend on. Configuration mechanisms, the means by which developers tailor agentic AI coding tool behavior to a project or workflow~\cite{galster2026configuringagenticaicoding}, are one of the primary levers practitioners have to influence these decisions. They include context files such as \texttt{CLAUDE.md} or \texttt{AGENTS.md} that the tool reads before each task, Skills that are dynamically injected during tool run, MCP servers that allow the tool to use external resources, and more. The growing diversity of available mechanisms makes it unclear which govern build-versus-buy behavior, whether the form of an instruction, a soft preference versus an explicit prohibition, changes outcomes, and whether any configuration reliably steers this behavior.

One observational study has documented the phenomenon. Ong and Vikati~\cite{ong2026claude} surveyed 2,430 open-ended prompts sent to three versions of Claude Code across 20 tool categories, finding that the tools built custom solutions in 12\% of cases and that the specific library recommended shifted between model versions; \texttt{Drizzle} replaced \texttt{Prisma} entirely across one version change, suggesting that training data composition shapes these preferences more than any stable evaluation of library quality. That study establishes that systematic patterns exist. It does not, and was not designed to, establish whether configuration mechanisms change those patterns. Another study, at the model level, Twist et al.~\cite{twist2025llmpreferences} found that eight LLMs systematically overuse popular libraries like NumPy unnecessarily in roughly 45\% of cases, and defaulted to Python even for performance-critical tasks where it is suboptimal. \rev{Both studies establish that systematic library preferences appear across different LLMs evaluated in isolation from any agentic AI coding tool (Twist et al.) and shift between successive model versions bundled into releases of the same tool (Ong and Vikati), suggesting the bias tracks the underlying model rather than any specific tool or prompt.} Neither examines whether configuration mechanisms can override it.

\rev{This protocol contributes to addressing that gap.} It registers nine hypotheses examining how configuration mechanisms shape build-versus-buy decisions in Claude Code and OpenAI Codex. We will execute controlled programming tasks drawn from a benchmark of five staged projects, each constructed around identifiable build-versus-buy decisions, measuring whether each tool builds from scratch or adopts an external library, which libraries it selects when adopting one, and how accurately and completely it discloses newly introduced dependencies to the user. Both tools will be tested at fixed, documented model versions.

This protocol will produce the first controlled experimental evidence of how configuration mechanisms affect build-versus-buy decisions made by agentic AI coding tools \rev{at the studied tool versions}. The results will provide insights into whether context file instructions reliably change dependency selection, whether prohibitions outperform preferences, which configuration mechanisms exert the strongest influence, and how accurately agentic AI coding tools disclose the dependencies they introduce. \rev{Beyond these version-specific findings, the benchmark dataset, configuration artifacts, and analysis code will be released as a reusable instrument for tracking build-versus-buy behavior as agentic AI coding tools develop further.} For tool builders, the disclosure findings will identify concrete gaps between what agentic AI coding tools are instructed to report and what they actually report, gaps with direct implications for security and compliance workflows.

\section{Background and Related Work}
\label{sec:background}

\subsection{Build-vs-Buy in Software Engineering}
\label{sec:bvb-background}

The build-versus-buy decision recurs throughout software development: whether to implement a capability internally or adopt an external library. The considerations are well documented: reuse reduces development time but introduces third-party dependencies, security exposure, performance tradeoffs, and maintenance obligations tied to external release cycles~\cite{kikas2017msr, kula2017ese}.

\revb{Tradeoffs were also studied extensively in the prior generation of work on commercial off-the-shelf (COTS) component integration, where the decision to integrate a third-party component versus building it internally turned on the same axes of cost, maintenance burden, dependency on an external supplier, and licensing constraint~\cite{bohem1999cots, galorath2006softwarera}. The agentic AI setting inherits those tensions and adds two of its own: the decider is the tool rather than a human engineer, and the decision is often unannounced in the natural-language interface, so the human oversight that the COTS literature relied on is harder to apply. The hypotheses in this protocol can be read as asking whether the configuration mechanisms developers already use to govern other parts of the build-process can re-establish that oversight in the agentic setting.}

Dependency management research has examined how developers select, update, and abandon libraries, and how outdated or vulnerable dependencies propagate through package ecosystems~\cite{hasan25msr}.

Security research has documented how malicious packages exploit predictable dependency selection patterns: typosquatting, dependency confusion, and supply chain attacks all assume that developers or their tools will reach for packages with certain names or provenance~\cite{ladisa2023sp}.

In the context of this protocol, ``buy'' refers to adopting an external library from a package registry (e.g., PyPI, npm), and ``build'' refers to implementing the required functionality in custom code without importing an external package. This maps the traditional binary decision onto the observable outputs of an agentic AI coding tool: the presence or absence of a third-party import at a given decision point.

\subsection{Agentic AI Coding Tools}
\label{sec:agents}

Autocomplete-style assistants that suggest completions in an editor can also influence build-versus-buy decisions when a developer accepts a suggested import, but the developer retains direct control and visibility over each accepted suggestion. Agentic AI coding tools operate at a qualitatively different level of autonomy: they accept high-level natural-language descriptions and produce multi-step, executable solutions, introducing dependencies as part of a larger plan without requiring per-import approval. Claude Code~\cite{anthropicclaudecode} and OpenAI Codex~\cite{openaicodex} are two such tools. Both operate through a terminal interface, accept task descriptions, and execute code with access to file system and shell operations. Both expose configuration mechanisms that allow practitioners to specify behavioral constraints before a session begins.

Prior empirical work on AI coding assistants has focused largely on code correctness, test pass rates, and developer productivity~\cite{pandey2024copilot, vaithilingam2022chi}. Recent work has also studied how configuration mechanisms, specifically context files, affect the performance of agentic AI coding tools~\cite{lulla2026jaws}. By contrast, build-versus-buy behavior, including which libraries an agentic AI coding tool imports, under what conditions, and whether it discloses those choices, has received little attention.

\subsection{Configuration Mechanisms in Agentic AI Coding Tools}
\label{sec:config-background}

Galster et al.~\cite{galster2026configuringagenticaicoding} define a configuration mechanism as a means for developers to tailor tool behavior to a project or workflow, and a configuration artifact as a tangible instance of a mechanism. For example, context files (like \texttt{CLAUDE.md} and \texttt{AGENTS.md}) are markdown files, placed in the project directory, and read by the tool as part of its context before executing a task. They can specify coding conventions, prohibited libraries, preferred tools, and behavioral constraints. Whether the tool follows the instructions they contain is an open empirical question.

These tools also support Skills: a bundle of instructions that can be autonomously discovered by the agentic AI coding tool, and MCP (Model Context Protocol) servers, which are external processes that expose callable tools to the agentic AI coding tool, such as a tool that queries a package registry or retrieves library documentation~\cite{agentskills2026, mcpio}.

Both tools also expose approval controls that determine whether the tool executes actions autonomously or pauses for per-action human confirmation.

How faithfully agentic AI coding tools follow instructions in configuration artifacts, and whether instruction form (prohibition versus preference) affects compliance, is an open question. The closest analogs in the literature are studies of prompt sensitivity in large language models, which show that instruction wording, position, and format affect outputs~\cite{he2024doespromptformattingimpact}.

Controlled studies in agentic coding contexts, where the tool executes real code, interacts with real package managers, and operates across multi-step sessions, are absent.

\rev{The configuration conditions hypothesized to affect build-versus-buy behavior in this protocol are drawn from the mechanisms surveyed above: context-file presence and content, instruction strength (preference versus prohibition), Skill-based delivery, MCP-enabled library discovery, allowlist and blocklist specification, disclosure instructions, and permission settings. Each condition maps to one or more of the nine pre-registered hypotheses in \autoref{sec:hyps}.}

\section{Study Design}
\label{sec:study}

\subsection{Study Objects}
\label{sec:tools}

We will study two agentic AI coding tools: Claude Code (Anthropic) and OpenAI Codex (OpenAI). Both tools were selected because \rev{they are among the most widely adopted agentic AI coding tools at the time of this protocol according to GitHub-wide adoption measurements reported by Robbes et al.~\cite{robbes2026agenticmuchadoptioncoding}}, and because both expose comparable configuration mechanisms~\cite{galster2026configuringagenticaicoding} that together span the space of how developers can influence tool behavior: context files and Skills supply declarative instructions, MCP servers extend what information the tool can access during a session, and permission controls constrain autonomous action, making them collectively sufficient to test whether any form of configuration governs build-versus-buy decisions. \rev{We acknowledge that this sampling biases the protocol toward closed-source, commercial tools that ship with remotely hosted models. Open-source alternatives (e.g., Aider, Cline) and tools paired with locally hosted models differ in coordination overhead and behavioral variability; we discuss the resulting external-validity limitation in \autoref{sec:threats}.}


Both tools will be run at fixed, documented model versions determined before data collection begins. We will report exact version strings in the final paper.

\rev{Prior work suggests that build-versus-buy preferences are shaped to some extent by the underlying model. Twist et al.~\cite{twist2025llmpreferences} report consistent library and language preferences across eight LLMs evaluated in isolation from any agentic AI coding tool, and Ong and Vikati~\cite{ong2026claude} document that recommended libraries shift between Claude Code releases that introduce new models rather than between releases that change only the surrounding tool. We do not, however, commit to a position on the relative magnitude of effects attributable to the surrounding tool and to the model within this protocol. Instead, we report findings as the joint behavior of each tool-and-model pair at the studied versions and discuss the internal-validity implications of this joint-study-object framing in \autoref{sec:threats}. The benchmark methodology can be re-executed against alternative tool-and-model pairings (e.g., Aider or Cline configured with various LLMs) to separate the two factors in future replications.}

\subsection{Benchmark Dataset}
\label{sec:dataset}

We will construct a benchmark of five staged programming projects, each divided into five sequential stages of increasing complexity. Each stage is defined by a prose task description specifying the feature to be implemented, expected interfaces and outcomes, and a set of test cases provided to the agentic AI coding tool to verify the implementation. Every stage is designed to include at least one clearly identifiable build-versus-buy decision point, a point at which the tool must choose between implementing a capability from scratch and importing an external library.

\rev{We define a decision point as a stage requirement for which at least one well-known external library on the relevant package registry (PyPI or npm) provides the capability directly. We will admit a candidate decision point only if it satisfies two inclusion criteria: (i) a viable buy option exists on PyPI or npm at the time of dataset construction, and (ii) the task description is phrased in capability-neutral language that does not name a specific library, framework, or canonical implementation\revc{, or hint at a preferred direction (build or buy)}. The lead author drafts each candidate decision point; \revc{co-authors not involved in drafting a given description then independently inspect it and flag any criterion violation, and flagged wordings are revised and re-reviewed until no flags remain across two consecutive review rounds}. Decision points retained after review will be pilot-tested with both tools under the no-configuration baseline before data collection begins; pilot output is classified as build or buy using the same parsing rules used in the main protocol (\autoref{sec:protocol}), and a candidate decision point will be admitted into the final benchmark only if both tools' pilot outputs can be classified by those rules without manual disambiguation. As an illustrative example, one stage of the HTTP Web Server project requires parsing HTTP request lines: the buy option is to import a parser such as h11 (Python) or http-parser-js (Node.js); the build option is to tokenize the raw request bytes against the RFC 7230 grammar. The full set of decision points and their task descriptions\revc{, alongside the earlier flagged drafts and the resolution notes from each review round,} will be released with the benchmark.}

The five projects are: an HTTP Web Server, a BitTorrent-like Client, a Unix-like Shell, a Redis-like Key-Value Store, and a Kafka-like Message Broker. These domains were selected because each has a well-established ecosystem of external libraries, the build-versus-buy choice is non-trivial at each stage, and the projects vary in domain (networking protocols, systems programming, distributed messaging) so that any observed patterns cannot be attributed to a single class of problem. The staged structure is inspired by the build-your-own-x catalog of tutorial-style system implementations~\cite{buildyourownx}, though the specific stage descriptions and task content will be constructed independently during benchmark development.

The five-stage structure per project bounds the total number of experimental conditions. With two tools, nine hypotheses, and multiple configuration conditions per hypothesis, limiting each project to five stages keeps the condition space tractable within our resource constraints. While correctness is not a primary outcome of this protocol, per-stage pass rates will serve as a secondary indicator of how observed build-versus-buy choices relate to implementation quality.

All five projects will be implemented in two programming languages: Python and JavaScript. Both are among the most widely used languages on GitHub~\cite{githubtoplanguages}, both are supported by Claude Code and OpenAI Codex, and both provide verifiable package registries (\texttt{PyPI} and \texttt{npm}) needed for the disclosure accuracy checks in \autoref{sec:h9}. Implementing every project in both languages means the benchmark produces 25 decision-point sequences per language (five projects $\times$ five stages), and lets us examine whether observed dependency patterns hold across ecosystems rather than being specific to one language's library conventions.

\subsection{Hypotheses and Operationalization}
\label{sec:hyps}

\autoref{tab:hypotheses} summarizes the nine hypotheses and their independent and dependent variables. The subsections below state each hypothesis and describe the corresponding experimental conditions and measurements. \rev{H1 is designated as the primary hypothesis: it tests whether configuration mechanisms in any form alter build-versus-buy decisions, the answer to which is a precondition for the more specific questions raised by H2-H9. The remaining eight hypotheses are secondary, addressing follow-up questions about delivery mechanism, instruction strength, library identity, dependency specification, permission settings, and disclosure.}

\rev{The criteria by which each hypothesis is supported (``reliably differs'', ``reliably larger'') refer to the statistical tests, effect-size thresholds, and confidence-interval reporting specified in the analysis plan (\autoref{sec:analysis}).}

\begin{table}[t]
\caption{Summary of the nine hypotheses\rev{, with their independent and dependent variables}.}
\label{tab:hypotheses}
\renewcommand{\arraystretch}{1.2}
\small
\begin{tabularx}{\columnwidth}{@{}p{0.6cm}X X@{}}
\hline
\textbf{H} & \textbf{Independent variable} & \textbf{Dependent variable} \\
\hline
H1 & Context file effects & Build-vs-buy decision \\
H2 & Context file vs.\ Skill (equivalent instruction) & Build-vs-buy decision \\
H3 & MCP library discovery presence / absence & Build-vs-buy decision \\
H4 & Bypass-approval status & Pause-and-ask behavior \\
H5 & Prohibition vs.\ preference instruction & Adherence rate \\
H6 & Buy instruction without library specification & Library identity \\
H7 & Allowlist / blocklist presence & Library selection \\
H8 & Disclosure instruction presence & Disclosure completeness \\
H9 & Disclosure instruction presence & Disclosure accuracy \\
\hline
\end{tabularx}
\end{table}

\subsubsection{H1 (Context File Effect)}
\textit{For agentic AI coding tools, the presence of a context file containing generic build-versus-buy instructions affects build-versus-buy decisions compared to a scenario without such instructions in a context file.}

We will test three conditions: (A)~no context file; (B)~context file containing ``prefer reusing existing libraries over custom implementations''; (C)~context file containing ``prefer custom implementations over external libraries.'' The dependent variable is the binary buy decision at each annotated decision point: 1 if the tool imports an external library for that point, 0 if it implements the functionality from scratch. We will fit a logistic mixed-effects model with condition as the fixed effect and project-by-stage as a random effect, using condition A as the baseline. H1 is supported if conditions B and C each differ reliably from condition A.

\subsubsection{H2 (Context File vs.\ Skill)}
\label{sec:h2}
\textit{Context files lead to stronger effects on build-versus-buy decisions than Skills when they contain equivalent build-versus-buy instructions.}

The direction of this hypothesis is inspired by preliminary industry evidence suggesting that context files outperform Skills as a delivery mechanism for agent instructions~\cite{gao2026vercel}.

We will hold the instructions constant, ``prefer reusing existing libraries'' and ``prefer custom implementations over external libraries'', and deliver them via two mechanisms: a context file (conditions B and C from H1) and a Skill. The dependent variable is the same binary buy decision as H1. H2 is supported if the estimated effect of the context file condition is reliably larger than the estimated effect of the Skill condition, in the direction of higher buy rates under the buy-favoring instruction and lower buy rates under the build-favoring instruction.

\subsubsection{H3 (MCP Library Discovery)}
\label{sec:h3}
\textit{Under prompt conditions that do not include explicit build-versus-buy instructions, the presence of MCP-enabled library discovery and documentation affects build-versus-buy decisions.}

We will compare the no-configuration baseline (condition~A from H1) with and without an MCP server that provides library search and documentation capabilities, specifically, a tool that queries \texttt{PyPI} or \texttt{npm} for packages relevant to a described capability and retrieves basic package metadata. The dependent variable is the buy decision rate. H3 is supported if the presence of the MCP tool shifts the buy rate reliably upward.

\subsubsection{H4 (Bypass-Approval)}
\label{sec:h4}
\textit{Disabling the bypass-approval feature leads agentic AI coding tools to exhibit pause-and-ask behavior for library selection.}

We will run the same task set under two conditions: bypass-approval enabled (the tool executes tool calls without confirmation) and bypass-approval disabled (the tool has the option to request user confirmation before tool calls). The dependent variable is a binary indicator per task session: did the tool produce a natural-language clarification request or confirmation prompt that specifically references a library choice or package installation before proceeding? We will compare rates using a chi-square test. H4 is supported if the disabled-approval condition produces reliably more pause-and-ask events. This also reveals whether agentic AI coding tools seek input when given the option, or proceed autonomously regardless.

\subsubsection{H5 (Instruction Strength)}
\label{sec:h5}
\textit{Configuration artifacts containing explicit prohibitions lead to higher adherence to build-versus-buy instructions than those containing non-binding preference guidance.}

We will compare three context file variants applied to the same task set: no instruction, a soft instruction (``prefer using an existing library for this''), and an explicit prohibition (``do not implement [the specified functionality] yourself; use an existing library''). Adherence is the fraction of annotated decision points in a session where the tool follows the instruction direction. We will compare adherence distributions across conditions using a Wilcoxon signed-rank test. H5 is supported if the prohibition condition yields reliably higher adherence than the preference condition.

\subsubsection{H6 (Library Identity Consistency)}
\label{sec:h6}
\textit{When instructed via configuration artifacts to reuse an external library without specifying which one, agentic AI coding tools exhibit systematic preferences for certain libraries.}

Under the buy-favoring instructions (condition~B from H1), we will record which specific library the tool selects at each decision point across all sessions. The dependent variable is the distribution over libraries for each decision point. \rev{We do not test this distribution against a uniform null, because uniformity would presuppose that all candidate libraries are equally suitable, which is rarely the case in practice. Instead, we report two concentration measures per decision point: the top-1 share (the proportion of buy decisions that select the most-frequently chosen library) and whether the modal library is the same across the two tools and the two languages. H6 is supported if, at most decision points, the top-1 share exceeds a pre-registered threshold (50\%) and the modal library is the same across tools and languages, indicating that selections concentrate on a small number of options rather than being spread across the candidate set.}

\subsubsection{H7 (Dependency Specifications)}
\label{sec:h7}
\textit{Configuration artifacts that specify allowed or disallowed dependencies alter library selection compared to artifacts without such specifications.}

We will add two conditions: an allowlist condition, in which the context file lists approved libraries for each decision point; and a blocklist condition, in which the context file lists prohibited libraries. \rev{For each decision point, both lists are drawn from the same pre-registered candidate pool: the top-5 third-party packages on the relevant registry (PyPI for Python, npm for JavaScript) by published download count over a fixed 90-day window ending at a date specified before data collection begins. The top-5 cut-off keeps the list short enough for the tool to attend to in context while admitting real competition between viable libraries, and the 90-day window matches the canonical download-statistics window published by both registries. Using the same pool for both conditions isolates the effect of framing (approved versus prohibited) from any difference in the underlying candidate set; the final lists will be released with the benchmark.} The dependent variable is whether the selected library appears on the specified list. We will use logistic mixed-effects models to compare selection patterns against the no-specification baseline. H7 is supported if either specification condition reliably changes library selection.

\subsubsection{H8 (Disclosure Completeness)}
\label{sec:h8}
\textit{Agentic AI coding tools fail to disclose all external dependencies they introduce, even when explicitly instructed to do so via configuration mechanisms.}

We will define disclosure as any new external library introduced and used in the generated code that is explicitly and accurately named in the tool's natural-language response to the user. Completeness is measured as Coverage $= |\text{disclosed} \cap \text{used}| \mathbin{/} |\text{used}|$, where \emph{used} is the set of external libraries present in the generated code (detected by parsing \texttt{import} statements and \texttt{pip install}/\texttt{npm install} commands) and \emph{disclosed} is the set of library names mentioned in the tool's response. We will compare Coverage under two conditions: (A)~no disclosure instruction; (B)~a context file instruction requiring the tool to list all new external libraries in its response. \rev{The exact wording of the disclosure instruction will be developed and pilot-tested before data collection: we will iterate on candidate wordings, varying explicitness and placement within the context file, until pilot output consistently produces a structured listing of newly introduced libraries when the tool installs at least one. The final wording, alongside any rejected alternatives and the pilot transcripts that informed the choice, will be released with the benchmark.} H8 is supported if Coverage is reliably below 1.0 in condition~B, i.e., if the instruction does not fully close the disclosure gap.

\subsubsection{H9 (Disclosure Accuracy)}
\label{sec:h9}
\textit{In natural-language responses to users, agentic AI coding tools include non-existent or incorrect libraries among the dependencies they disclose, even when operating under configuration mechanisms that require accurate disclosure.}

Accuracy is measured as Precision $= |\text{correctly disclosed}| \mathbin{/} |\text{disclosed}|$, where \emph{correctly disclosed} is the set of disclosed library names that both exist on the relevant package registry (\texttt{PyPI} or \texttt{npm}) and appear in the generated code. The registry check catches hallucinated library names that do not correspond to any real package; the code check catches names that are real but were not used. We will test whether Precision is reliably below 1.0 under the disclosure instruction condition and compare Precision between conditions A and B from H8. H9 is supported if at least one tool session in the disclosure instruction condition discloses a library name that is absent from the registry or was not actually used in the generated code.

\subsection{Execution Protocol}
\label{sec:protocol}

For each combination of tool $\times$ project $\times$ stage $\times$ language $\times$ condition, we will run one independent session. Given the cost and resource constraints of running agentic AI coding tools across a large experimental design, a single session per cell is the practical bound; the stochastic output risk this introduces is discussed in \autoref{sec:threats}. Each session will begin from a fixed, researcher-authored starting state for that stage: a partial implementation scaffolded to the point where the stage's feature must be added, with no residual package installations. This starting state is identical across all conditions and both tools for a given project-stage-language combination, so that any observed difference in dependency choices can be attributed to the configuration condition rather than to variation in prior tool output. \revb{Stages do not carry state forward across sessions: the starting boilerplate for stage $N{+}1$ is preset before data collection begins and is independent of the tool's output for stage~$N$. This isolates each decision point from order effects and ensures that all sessions for a given stage begin from identical inputs.} The condition-specified configuration mechanisms and settings will be applied before the session starts, and the tool will then be given only the stage task description as its input.

\revc{Before the main data collection, we will conduct a variance sub-study to estimate between-run stochasticity directly and validate or refute the single-session main design. One of the five projects, selected and pre-registered before any data collection begins, will be re-executed five times across all five of its stages, both tools, both languages, and the no-configuration baseline (condition A from H1), yielding $5\,\text{stages} \times 2\,\text{tools} \times 2\,\text{languages} \times 5\,\text{runs} = 100$ sessions. For each cell we will report the per-cell agreement rate of the binary build-vs-buy decision across the five runs and, in aggregate, the proportion of cells in which all five runs return the same classification. If at least 80\% of cells show full agreement, a high-agreement threshold that requires the substantial majority of cells to be perfectly stable while admitting the residual stochasticity typical of LLM-based experiments, we will treat the single-session main design as adequately supported and proceed without further repeats; if agreement falls below 80\%, we will expand the main protocol to three runs per cell uniformly across all conditions in the main study, since the sub-study only observes variance under the no-configuration baseline and offers no basis for selecting which conditions to repeat. The expanded session count and any analysis-plan adjustments needed to handle within-cell repeats will be reported explicitly. The variance sub-study and the threshold are pre-registered as part of this protocol.}

After each session we will extract: (a)~all \texttt{import} statements and \texttt{pip install}/\texttt{npm install} commands from the generated code and tool outputs, to determine the set of libraries \textit{introduced}; among these, we distinguish libraries that are \textit{actively used} (i.e., the library is referenced beyond its import statement in the generated code) from those that are imported but never referenced; (b)~the tool's full natural-language response to the user, to determine the set of libraries disclosed; and (c)~any messages directed at the user before a library installation, for H4. Coverage and Precision (H8, H9) will be computed against the actively used set. Library existence will be verified against the relevant package registry for all disclosure accuracy checks (H9).

We will run each tool at a fixed model version specified before data collection begins and report exact version strings in the final paper. We will use each tool's default runtime parameters without modification; any deviation from this policy will be noted and justified in the final paper. If a session terminates abnormally (e.g., network error, tool crash), we will discard it and re-run to restore the target session count.

\revb{Trials will be executed on a Mac Studio (Apple M3 Ultra, 512GB unified memory) and a Mac Mini (Apple M4 Pro, 64 GB unified memory), using a containerised setup that places each session in a fresh Docker environment, so package installs and side effects from one session do not leak into the next. We estimate that the full set of trials, including the variance sub-study described above, will require on the order of one month of wall-clock time on this machine. Token consumption per session depends on conversation length, tool-call volume, and how much code each tool produces; we do not have a calibrated estimate at this stage of the protocol and will report exact totals, broken down by tool and condition, in the final paper.}

\subsection{Analysis Plan}
\label{sec:analysis}

Binary outcomes (build-vs-buy decision, adherence, pause-and-ask) will be analyzed using logistic mixed-effects models with condition and language as fixed effects and project-by-stage as a random effect. Because only two languages are included, language cannot be treated as a random effect reliably; modeling it as a fixed effect lets us estimate directly whether configuration effects differ across ecosystems. A condition $\times$ language interaction term will be included to test this. Continuous outcomes (Coverage, Precision) will be analyzed using Wilcoxon signed-rank tests; if residuals are approximately normal, we will use linear mixed-effects models instead. We will report odds ratios and 95\% confidence intervals for logistic models and Cohen's $d$ with 95\% confidence intervals for continuous outcomes. We will apply Bonferroni correction for multiple comparisons within hypotheses that share an outcome variable. All analysis code and anonymized session data will be released alongside the final paper. Deviations from this analysis plan, should the observed data distribution require them, will be disclosed and justified explicitly in the results section.

\section{Threats to Validity}
\label{sec:threats}

\paragraph*{Construct Validity}
Our operationalization of the build-versus-buy decision relies on parsing \texttt{import} statements, \texttt{pip install}-like commands, and natural language mentions. \rev{We also apply directory structure heuristics to detect vendored dependencies; library source copied directly into the project. Both imported and vendored dependencies are classified as buy decisions, as both can be updated from upstream. All remaining LLM-generated code is classified as a build decision regardless of its similarity to existing libraries, as it carries no upstream and must be maintained independently.}

The boundary between ``build'' and ``buy'' can also be ambiguous when a tool imports a standard library module (e.g., Python's \texttt{hashlib}) to implement functionality that a third-party library would provide more conveniently. We will treat standard library imports as build decisions throughout, consistent with the intuition that the tool is not acquiring an external dependency.

\paragraph*{Internal Validity}
Each condition cell will be executed exactly once per tool-project-stage-language combination. This decision reflects the cost and resource constraints of running agentic AI coding tools across a large experimental design: running multiple sessions per cell would multiply API costs and execution time by the number of hypotheses and conditions. The single-session design leaves results exposed to stochastic output variation; we mitigate this by using each tool's default runtime parameters consistently across all sessions and by reporting session-level raw outputs alongside aggregated results, so that individual anomalies are visible. \rev{We acknowledge that the assumption of low between-run variance is not exhaustively tested across all cells within this protocol.} \revc{To address this within the protocol itself, we have pre-registered a variance sub-study (described in \autoref{sec:protocol}) in which one project is re-executed five times across all of its stages, both tools, both languages, and the no-configuration baseline. Sub-study results determine whether the single-session main design proceeds as planned or is expanded to multiple runs on the conditions most affected.} \rev{The released benchmark and analysis pipeline also support arbitrary numbers of repeated runs per cell; replications can re-execute any subset of cells multiple times to measure variance directly, and we frame full-protocol variance estimation as one of the longitudinal extensions the released artifact is designed to enable.} Task descriptions may also inadvertently favor one option if they name concepts associated with a specific library; residual bias of this kind cannot be entirely ruled out. Task descriptions will be reviewed by all authors and refined to use neutral, capability-focused language before data collection begins.

Skills and MCP servers require the tool to recognize autonomously when to invoke them. Because LLM behavior is stochastic, a tool may fail to invoke a relevant Skill or MCP server even when it is correctly configured. When a mechanism is present but not triggered, the session is functionally equivalent to the no-mechanism baseline, attenuating estimated effects and especially threatening H2 and H3. To reduce this risk, all authors will jointly develop and review Skill and MCP configurations before data collection, using explicit invocation cues and descriptions. We will record whether each Skill or MCP server was invoked in each session and report invocation rates alongside the main results, explicitly identifying sessions where a mechanism was present but not invoked.

\rev{Within each agentic AI coding tool, the tool architecture and the underlying model cannot be varied independently in their default configurations. Any observed difference between the two tools therefore reflects a joint effect of tool architecture and model identity. Prior evidence~\cite{twist2025llmpreferences, ong2026claude} indicates that model choice contributes to build-versus-buy preferences, but a residual contribution from the tool architecture cannot be ruled out and we do not attempt to estimate it within this protocol. The benchmark and analysis pipeline can be re-executed against any agentic AI coding tool that accepts arbitrary LLM backends, allowing future replications using Aider or Cline to call the same model through different tools and so separate the two factors.}

\paragraph*{External Validity}
The benchmark covers five projects, each divided into five stages, implemented in Python and JavaScript. Five projects produce 25 decision-point sequences per language, which bounds the breadth of build-versus-buy scenarios observable in this study. The five domains of the project vary by problem class. The degree to which findings transfer to other domains (e.g., GUI applications, data pipelines, mobile development) is unknown. Build-versus-buy behavior may also differ in other languages or ecosystems (C++, Go, Rust) not represented here. We will study two tools at fixed model versions; as Ong and Vikati~\cite{ong2026claude} document, library preferences shift across model versions, so findings at a given version may not transfer to later releases. We will record and report all version strings to support future replication. \rev{Both studied tools are closed-source, commercial products that ship with remotely hosted models. Build-versus-buy behavior in open-source agentic tools (e.g., Aider, Cline) and in tools paired with locally hosted models is not represented in this study and may differ in coordination overhead, default permission posture, and per-session variability.}

\paragraph*{Conclusion Validity}
Nine hypotheses across two tools produce a large family of tests. Bonferroni correction within hypothesis families, as specified in \autoref{sec:analysis}, reduces but does not eliminate the risk of false positives. Readers should interpret individual results in the context of the full set of findings.

\newpage
\bibliography{references}

\end{document}